\documentclass[amsmath, 12pt]{article}
\usepackage{amsfonts}
\usepackage{amscd,amssymb,amsthm,amsmath}
\usepackage{euscript}
\usepackage{mathrsfs}
\usepackage{euscript}
\usepackage{esvect}
\usepackage{comment}
\usepackage{xcolor}
\usepackage{hyperref}
\usepackage{lmodern}
\usepackage{bm}
\usepackage{graphicx}

\newtheorem{theorem}{Theorem}

\newtheorem{lemma}{Lemma}

\newtheorem{example}{Exemple}
\newtheorem{remark}{Remark}

\begin{document}

\title{On the Construction of New Toric Quantum Codes and Quantum Burst-Error Correcting Codes}

\author{Cibele Cristina Trinca\thanks{The author is with the Department of Biotechnology and Bioprocess Engineering, Federal University of Tocantins, Gurupi , TO, Brazil (e-mail: cibtrinca@yahoo.com.br).}, J. Carmelo Interlando\thanks{The author is with the Department of Mathematics and Statistics, San Diego State University, San Diego, CA, USA (e-mail: carmelo.interlando@sdsu.edu).}, \\ Reginaldo Palazzo Jr.\thanks{The author is with the School of Electrical and Computer Engineering, State University of Campinas, Brazil (e-mail: palazzo@dt.fee.unicamp.br).}, Antonio Aparecido de Andrade\thanks{The author is with the Department of Mathematics, S\~{a}o Paulo State University, Brazil (e-mail: antonio.andrade@unesp.br).} and \\ Ricardo Augusto Watanabe\thanks{The author is with the School of Mathematics, Statistics and Scientific Computing, State University of Campinas, Brazil (e-mail: ricardoaw18@gmail.com).}}

\date{\today}
\maketitle

\begin{abstract}
\noindent A toric quantum error-correcting code construction procedure is presented in this work. A new class of an infinite family of toric quantum codes is provided by constructing a classical cyclic code on the square lattice $\mathbb{Z}_{q}\times \mathbb{Z}_{q}$ for all odd integers $q\geq 5$ and, consequently, new toric quantum codes are constructed on such square lattices regardless of whether $q$ can be represented as a sum of two squares. Furthermore this work supplies for each $q$ the polyomino shapes that tessellate the corresponding square lattices and, consequently, tile the lattice $\mathbb{Z}^{2}$. The channel without memory to be considered for these constructed toric quantum codes is symmetric, since the $\mathbb{Z}^{2}$-lattice is autodual. Moreover, we propose a quantum interleaving technique by using the constructed toric quantum codes which shows that the code rate and the coding gain of the interleaved toric quantum codes are better than the code rate and the coding gain of Kitaev's toric quantum codes for $q=2n+1$, where $n\geq 2$, and of an infinite class of Bombin and Martin-Delgado's toric quantum codes. In addition to the proposed  quantum interleaving technique improves such parameters, it can be used for burst-error correction in errors which are located, quantum data stored and quantum channels with memory.     
\end{abstract}



\section{Introduction}
There are currently several experimental proposals for building a quantum computer, however, the work is faced with several challenges, one being the fragility of quantum  information: Interactions of the system with its environment may destroy the superposition states, causing information loss. This event is called (quantum) decoherence.  The problem can be circumvented by isolating the system from its environment, which in turn may be difficult to achieve. One way to overcome such an issue is by using quantum error-correcting codes by encoding quantum states so that they are resistant to noise. Decoding is performed at the time when the states must be recovered.  

Constructions of quantum error-correcting codes are strongly based on their classical counterparts; however, there are some fundamental differences between classical and quantum information, e.g., (i) copying a qubit is physically impossible \cite{Kitaev1} and (ii) measurements destroy the quantum state in most cases, which in turn prevent its recovery \cite{Kitaev1}.  

Within the class of stabilizer quantum error-correcting codes, Kitaev introduced an alternative language using topology \cite{Kitaev1}. His proposal was to use certain properties of particles confined to a plane to perform topological quantum computation. That terminology comes from the fact that those properties are related to the topology of the physical system. Continuous deformations caused by the environment can alter those properties, and as consequence, we would naturally have quantum computation resistant to errors.

Kitaev started that investigation via toric codes \cite{Kitaev2}. To construct them, qubits are associated to the edges of a square lattice on the torus; hence,  the total number of edges is equal to the length of the code. The stabilizer operators are related to the vertices and the faces of the square lattice, and the coded qubits are determined according to the genus of the surface. Lastly, the distance of the code is determined by the homology group of the surface. Toric codes can be generalized to a class known as topological quantum codes by considering other two-dimensional orientable surfaces different from the torus.

In \cite{BombinDelgado}, Bombin and Martin-Delgado considered a new version of Kitaev's code, introducing a class of quantum codes with distance 3 obtained from embeddings of complete graphs in surfaces with corresponding genera. Remarkably, these codes have higher rates of information than the original ones.
Furthermore, the version of toric codes introduced by Bombin and Martin-Delgado provides new insight into quantum toric codes, relating them to a topic that has been extensively studied within the context of classical codes, namely, lattice codes \cite{Conway}. From this connection, it is possible to use group theory and combinatorics to obtain several new quantum toric codes.

Let $q=2n+1$, where $n\geq 2$ is an integer. In \cite{ClariceArtigo}, the quadratic form $x^2+y^2$ which is associated with the lattice $\mathbb{Z}^{2}$ is used to construct classical codes and, consequently, toric quantum codes on the square lattice $\mathbb{Z}_{q}\times \mathbb{Z}_{q}$ in the particular case where $q$ can be represented as a sum of two squares. On the other hand, in our contribution, via a combinatorial method, a classical code on the square lattice $\mathbb{Z}_{q}\times \mathbb{Z}_{q}$ is constructed for all odd integers $q$ ($q\geq 5$) and, consequently, we obtain new toric quantum codes on the square lattices $\mathbb{Z}_{q}\times \mathbb{Z}_{q}$ for all odd integers $q$ ($q\geq 5$). Therefore, new toric quantum codes are obtained regardless of whether $q$ can be represented as a sum of two squares. The parameters of the new toric quantum codes being proposed are given by $[[2q,2,d_{M}=3]]$ with $q=2n+1$ and $n=2,3,4$, and $[[2q,2,d_{M}=4]]$ with $q=2n+1$ and $n\geq 5$, where $d_M$ denotes the Mannheim distance. Nonetheless, the duality of the lattice on which the toric quantum code is constructed influences the error correction pattern \cite{LivroBombin}, that is, since the $\mathbb{Z}^{2}$-lattice is autodual, then the channel without memory is symmetric.     

In classical information theory there are mainly two types of error models: the independent noise model proposed by Shannon and the adversarial noise model considered by Hamming. Errors in these two models are usually called random errors and burst errors. Correspondingly, there are random error-correcting codes and burst error-correcting codes to deal with these two different types of errors \cite{Lin}. Actually channels tend to introduce errors which are located in a short interval, i.e., the burst errors. These errors could be commonly found in communication systems and storage mediums as a result of a stroke of lightning in wireless
channels or scratch on a storage disc.   

In the quantum regime, quantum errors can be independent or correlated in space and time. Hence there are counterparts of quantum random error-correcting codes \cite{CaldShor,Steane1,Wilde} and quantum burst error-correcting codes \cite{Fan,Vatan,Kawabata,Fan1}. Analogously to the classical case, quantum channels usually have memory \cite{Werner} or introduce errors which are located \cite{Caruso}, that is, quantum burst errors. 

The construction and investigation of quantum burst error-correcting codes have received far less attention compared to the development of standard quantum error-correcting codes or entanglement-assisted quantum error-correcting codes \cite{Wilde,Lai}. 

In quantum computing, quantum memory is the quantum-mechanical version of ordinary computer memory. Whereas ordinary memory stores information as binary states (represented by ``1"s and ``0"s), quantum memory stores a quantum state for later retrieval. These states hold useful computational information known as qubits. Unlike the classical memory of everyday computers, the states stored in quantum memory can be in a quantum superposition, giving much more practical flexibility in quantum algorithms than classical information storage.

Quantum memory is essential for the development of many devices in quantum information processing, including a synchronization tool that can match the various processes in a quantum computer, a quantum gate that maintains the identity of any state, and a mechanism for converting predetermined photons into on-demand photons. Quantum memory can be used in many aspects, such as quantum computing and quantum communication. Continuous research and experiments have enabled quantum memory to realize the storage of qubits \cite{Lvovsky}. 

Our contribution also presents a new interleaving technique by using the constructed toric quantum error-correcting codes on the square lattices $\mathbb{Z}_{q}\times \mathbb{Z}_{q}$ for all odd integers $q$ ($q\geq 5$). The motivation to use such a technique is related to providing quantum burst-error correcting codes which improve the code rate and the coding gain \cite{rates} and burst-error correction in errors which are located, quantum data stored and quantum channels with memory.     

The interleaving technique presented in this work shows that the code rate and the coding gain of the interleaved toric quantum codes are better than the code rate and the coding gain of Kitaev's toric quantum codes for $q=2n+1$, where $n\geq 2$, and of an infinite class of Bombin and Martin-Delgado's toric quantum codes.

This paper is organized as it follows. In section II we review the basic results of lattice and polyomino theory. In section III we review previous results of toric quantum codes. In section IV through a combinatorial method a classical code on the square lattice $\mathbb{Z}_{q}\times \mathbb{Z}_{q}$ is constructed for all odd integers $q$ ($q\geq 5$) and, consequently, we obtain new toric quantum codes on the square lattices $\mathbb{Z}_{q}\times \mathbb{Z}_{q}$ for all odd integers $q$ ($q\geq 5$). Therefore, new toric quantum codes are obtained regardless of whether $q$ can be represented as a sum of two squares. Furthermore section IV provides the polyomino shapes that can tessellate such a square lattice. Section V supplies a quantum  interleaving technique by using the constructed toric quantum codes to provide quantum burst-error correcting codes by interleaving such toric quantum codes.

\section{Lattices and Polyominoes}
The background material presented in this section can be found in \cite{ClariceArtigo} and \cite{ijam}. A large class of problems in coding theory is related to the properties of lattices \cite{ijam4,Agnaldo,ijam1,ijam2,ijam3}.


Let $\{ \bm a_1, \ldots, \bm a_n\}$ be a basis for the $n$-dimensional real Euclidean space, $\mathbb R^n$, where $n$ is a positive integer. An $n$-dimensional lattice $\Lambda$ is the set of all points of the form $u_1 \bm a_1 + \cdots + u_n \bm a_n$ with integral $u_1,\ldots, u_n$. Thus, $\Lambda$ is a discrete additive subgroup of $\mathbb R^n$. This property leads to the study of subgroups (sublattices) and coset decomposition (partitions). An algebraic way to obtain sublattices from lattices is via a scaling matrix $A$ with integral entries. Given a lattice $\Lambda$, a sublattice $\Lambda ^{\prime}=A \Lambda$ can be obtained by transforming each vector $\lambda \in \Lambda$ to $\lambda ^{\prime} \in \Lambda ^{\prime}$ according to $\lambda ^{\prime} = A \lambda$.

Every building block that fills the entire space with one lattice point in each region, when repeated many times, is called a \textit{fundamental region} of the lattice $\Lambda$. There are several ways to choose a fundamental region for a lattice $\Lambda$, however the volume of the fundamental region is uniquely determined by $\Lambda$ \cite{Conway}. Let $V(\Lambda)$ denote the volume of a fundamental region of the $n$-dimensional lattice $\Lambda$. For a sublattice $\Lambda ^{\prime} = A \Lambda$, we have that $\frac{V(A \Lambda)}{V(\Lambda)} = |\det A|$. Also, the number of cosets $q$ of $\Lambda'$ in $\Lambda$ is given by $q = |\det A|$ \cite{Clarice}.

Quadratic forms provide an alternative method of study for lattices which is especially useful to investigate their arithmetical properties \cite{Conway}. More specifically, let $G$ be a matrix whose rows form a basis for a lattice $\Lambda$. The quadratic form $\bm x GG^{tr} \bm x^{tr}$, where $tr$ denotes transpose, is said to be associated with $\Lambda$.  An important parameter of the lattice is its volume (more precisely, the volume of its fundamental region), which is given by the square root of the determinant of $GG^{tr}$. 
 
A polyomino is a plane geometric figure formed by joining one or more equal squares edge to edge \cite{polyominoGolomb}. It is a polyform whose cells are squares. It may be regarded as a finite subset of the regular square tiling. Such polyominoes with area $q$ (a composition of $q$ squares) will be used as fundamental regions to tessellate the lattice $\mathbb{Z}^{2} = \{(x,y) \ | \ x, y \in \mathbb Z \}$.

Polyominoes are examples of combinatorial geometry and are used in classical coding theory to obtain {\it close-packed codes}. A close-packed code corresponds to a tessellation of the group consisting of the cosets of $q\mathbb{Z}^{2}$ in $\mathbb{Z}^{2}$, which in turn is isomorphic to $\mathbb{Z}_{q}\times \mathbb{Z}_{q}$, by translations of a given polyomino shape \cite{Golomb}. This shape may be viewed as the decision region associated with each codeword of a standard error-correcting code.   

\section{Toric Quantum Codes}
The material presented in this section can be found in \cite{Clarice, ClariceArtigo, Kitaev1, Kitaev2}. A quantum error-correcting code (QEC) is the image of a linear mapping from the $2^{k}$-dimensional Hilbert space $H^{k}$ to the $2^{n}$-dimensional Hilbert space $H^{n}$, where $k<n$. The codewords are the vectors in the $2^{n}$-dimensional space. The \textit{minimum distance d} of a quantum error-correcting code $C$ is the minimum distance between any two distinct codewords, that is, the minimum Hamming weight of a nonzero codeword. A quantum error-correcting code $C$ of length $n$, dimension $k$ and minimum distance $d$ is denoted by $[[n,k,d]]$. A code with minimum distance $d$ is able to correct up to $t$ errors, where $t=\lfloor \frac{d-1}{2} \rfloor$ \cite{Shor}. 

A stabilizer code $C$ is the simultaneous eigenspace with eigenvalue 1 comprising all the elements of an Abelian subgroup $S$ of the Pauli group $P_{n}$, called the \textit{stabilizer group}. The elements of the Pauli group on $n$ qubits are given by 
\[
P_{n}=\{\pm I, \pm iI, \pm X, \pm iX, \pm Y, \pm iY, \pm Z, \pm iZ\}^{\otimes n}, \ \mathrm{where}
\]
\[
I=\left( \begin{array}{cc}
                    1 & 0 \\
                    0 & 1 \\
                    \end{array}
                    \right), \ X=\sigma_{x}=\left( \begin{array}{cc}
                    0 & 1 \\
                    1 & 0 \\
                    \end{array}
                    \right),
\]
\begin{equation}
Y=\sigma_{y}=\left( \begin{array}{cc}
                    0 & -i \\
                    i & 0 \\
                    \end{array}
                    \right) \ \mathrm{and} \ Z=\sigma_{z}=\left( \begin{array}{cc}
                    1 & 0 \\
                    0 & -1 \\
                    \end{array}
                    \right).
\end{equation}
Thus, $C=\{| \psi \rangle \in H^{n} \ | \ M | \psi \rangle = | \psi \rangle, \ \forall \ M\in S\}$ \cite{Gott}.

Kitaev's toric codes form a subclass of stabilizer codes and they are defined in a $q\times q$ square lattice of the torus (Figure 1). Qubits are in one-to-one correspondence with the edges of the square lattice. The parameters of this class of codes are $[[2q^{2},2,q]]$, where the code length $n$ equals the number of edges $|E|=2q^{2}$ of the square lattice. The number of encoded qubits is dependent on the genus of the orientable surface. In particular, codes constructed from orientable surfaces $gT$ (connected sum of $g$ tori $T$) encode $k=2g$ qubits. Thus, codes constructed from the torus, an orientable surface of genus 1, have $k=2$ encoded qubits. The distance is the minimum between the number of edges contained in the smallest homologically nontrivial cycle of the lattice and the number of edges contained in the smallest homologically nontrivial cycle of the dual lattice. Recall that the square lattice is self-dual and a homologically nontrivial cycle is a path of edges in the lattice which cannot be contracted to a face. Therefore the smallest of these two paths corresponds to the orthogonal axes either of the lattice or of the dual lattice. Consequently, $d=q$ \cite{DennisKitaev}.

\begin{figure}
    \centering
    \includegraphics[scale=0.35]{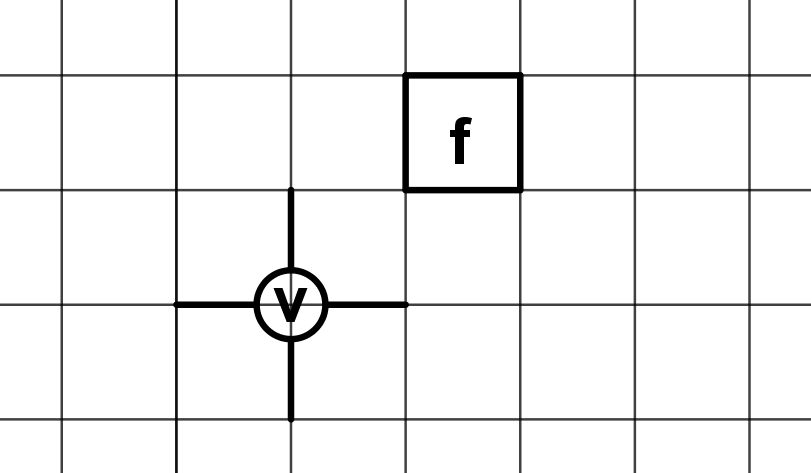}
    \caption{Square lattice of the torus, from \cite{ClariceArtigo}.}
    \label{fig:1}
\end{figure}

The stabilizer operators are associated with each vertex and each face of the square lattice (lattice) (Figure \ref{fig:1}). Given a vertex $v\in V$, the vertex operator $A_{v}$ is defined by the tensor product of $\sigma_{x}$ -- corresponding to each one of the four edges which have $v$ as a common vertex -- and the operator identity acting on the remaining qubits. Analogously, given a face $f\in F$, the face operator $B_{f}$ is defined by the tensor product $\sigma_{z}$ -- corresponding to each one of the four edges forming the boundary of the face $f$ -- and the operator identity acting on the remaining qubits. In particular, 
\begin{equation}    
A_{v}=\bigotimes_{j\in E} \sigma_{x}^{\delta (j\in E_{v})} \ \mathrm{and} \ B_{f}=\bigotimes_{j\in E} \sigma_{z}^{\delta (j\in E_{f})} \ \cite{ClariceArtigo},
\end{equation}
\noindent where $\delta$ is the Kronecker delta. 

The toric code consists of the space fixed by the operators $A_{v}$ and $B_{f}$ and it is given as 
\begin{equation}
C=\{| \psi \rangle \in H^{n} \ | \ A_{v} | \psi \rangle = | \psi \rangle \ \mathrm{and} \ B_{f} | \psi \rangle = | \psi \rangle , \ \forall \ v,f \}.
\end{equation}

The dimension of $C$ is 4, that is, $C$ encodes $k=2$ qubits. 

\section{Toric Quantum Codes: A Combinatorial Approach}\label{toricquantum}

Under the algebraic point of view, Kitaev's code can be characterized as the group consisting of the cosets of $q\mathbb{Z}^{2}$ in $\mathbb{Z}^{2}$, which in turn is isomorphic to $\mathbb{Z}_{q}\times \mathbb{Z}_{q}$.  The identifications of the opposite edges of the region delimited by $\mathbb{Z}_{q}\times \mathbb{Z}_{q}$ result in its identification with a flat torus; for the sake of simplicity, we call this region \textit{lattice} or \textit{$q\times q$ square lattice}. The area associated with the lattice $\mathbb{Z}_{q}\times \mathbb{Z}_{q}$ is $q^{2}$. Since each edge belongs simultaneously to two square faces of the lattice, there are $2q^{2}$ edges, that is, $n=2q^{2}$ qubits. The qubits to be encoded are related to the essential cycles of the surface and, in the case of the torus, there are two cycles (meridian and parallel); consequently, $k=2$. Similarly to the classical case, the minimum distance of the code is defined as the minimum between the least number of edges to be traversed in the lattice and the least number of edges to be traversed in the dual lattice with respect to a specific coset representative. Therefore this leads to $d=q$. 

The goal of this section is to construct toric quantum codes by using fundamental regions of sublattices of the lattice $\mathbb{Z}^{2}$. These fundamental regions are associated with polyominoes with area $q$. In fact, the area of each polyomino could be any value that divides the area $q^{2}$ of the lattice (or square lattice) $\mathbb{Z}_{q}\times \mathbb{Z}_{q}$. As an example, for $q=5$, Figure 2 shows two polyominoes with area $q=5$ associated with fundamental regions which tessellate the lattice (or square lattice) $\mathbb{Z}_{5}\times \mathbb{Z}_{5}$. 

 \begin{figure}[ht!]
    \centering
    \begin{minipage}{0.49\textwidth}
        \centering
        \includegraphics[scale=0.5]{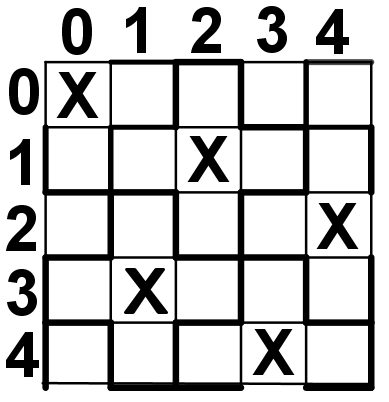} 
        \label{figura3Clarice_A}
    \end{minipage} 
    \begin{minipage}{0.49\textwidth}
        \centering
        \includegraphics[scale=0.5]{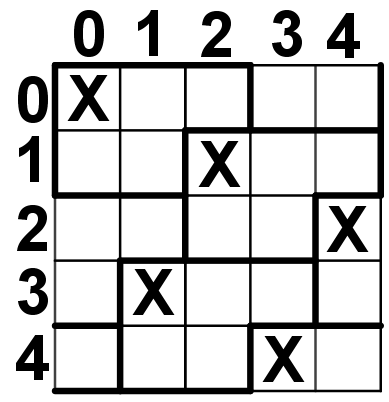} 
        \label{figura3Clarice_B}
    \end{minipage}
    \caption{Two representatives of regions with area 5, from \cite{Clarice}.}
\end{figure}

However, before determining the fundamental regions of the tessellation, the set of elements which are representatives of these regions and denoted in the figures by an $X$ mark must be known. In general, these representatives are identified as codewords of a cyclic code whose codewords have coordinates $(x,y)\in \mathbb{Z}_{q}\times \mathbb{Z}_{q}$ that indicate where the polyomino will be placed. For instance, in Figure 2 the representatives are codewords of a cyclic code whose coordinates are given by $(0,0), \ (2,1), \ (4,2), \ (1,3)$ and $(3,4)$. Then we are faced with the combinatorial problem of finding the possible shapes/polyominoes (decision regions) associated with each codeword of this cyclic code.  

This set of representatives corresponds to the set of codewords of a classical code, that is, a vector subspace of $\mathbb{Z}_{q}\times \mathbb{Z}_{q}$ denoted by $\mathcal{A}$. In order to obtain a polyomino with area $q$, the cardinality of $\mathcal{A}$, denoted by $|\mathcal{A}|$, must be equal to $q$. 

The quadratic form associated with the lattice $\mathbb{Z}^{2}$ is given by $x^{2}+y^{2}$. Henceforth, we shall say that the quadratic form associated with the square lattice $\mathbb{Z}_{q}\times \mathbb{Z}_{q}$ is given by $x^{2}+y^{2} = q$, see \cite{ClariceArtigo}.

In \cite{ClariceArtigo}, such a quadratic form is used to construct classical codes and, consequently, obtain toric quantum codes constructed on the square lattice $\mathbb{Z}_{q}\times \mathbb{Z}_{q}$. Consider the case where $|\mathcal{A}|=q$ and $x$ and $y$ are relatively prime numbers with $q=x^{2}+y^{2}$. Hence the group associated with $\mathcal{A}$ is the cyclic group generated by $\langle (x,y) \rangle$ (one generator). Now if $x$ and $y$ are not relatively prime numbers, that is, $\gcd(x,y) > 1$, then the group associated with $\mathcal{A}$ is $\langle (x,y),(-y,x) \rangle$, whose cardinality equals $q$. The operation addition modulo $q$ is used in these cases. The authors note that in the cases where $q=x^{2}+y^{2}$ with $\gcd(x,y)=1$, the corresponding set of representatives of the polyominoes (codewords) $\mathcal{A}$ is a perfect code, that is, $\mathcal{A}$ has only one representative (codeword) $X$ in each row or column \cite{Sueli}.  

Once the subspace given by the representatives (codewords) is known, it is possible to choose the polyominoes that may tessellate the square lattice. From this classical construction, the authors construct toric quantum codes; the toric quantum code associated with this tessellation may be established in the same way as it was in the Kitaev's code. The code length is given by the number of edges of the polyomino, that is, $n=2q$, since the polyomino has area $q$ and each edge belongs simultaneously to two square faces of the original square lattice. The code dimension is $k=2$, since this code is constructed in the torus. From the fact that the $\mathbb{Z}^{2}$-lattice is autodual, the code distance is defined as the minimum number of edges in the square lattice between two representatives of the polyominoes (codewords) \cite{LivroBombin}. Such a distance is given by $d_{M}=\min\{|x|+|y| \ | \ (x,y)\in \mathcal{A}\}$, where $w_{M}(x,y)=|x|+|y|$ is known as the \textit{Mannheim weight} of $(x,y)$. Consequently, the parameters of the toric quantum codes generated by these tessellations are $[[2q,2,d_{M}]]$. 

Although the polyomino used to tessellate the square lattice may have different shapes, the parameters $n, \ k$, and $d_{M}$ will be the same. However, the duality of the lattice on which toric quantum codes are constructed influences the error correction pattern \cite{LivroBombin}, that is, if the corresponding lattice is autodual, then the channel without memory is symmetric; but if the corresponding lattice is not autodual, then the channel without memory is not symmetric.    

\begin{theorem}\label{Genus} \cite{cohn} (Genus) For $q$ a prime integer, the equation $x^{2}+y^{2}=q$ can be solved for integers $x$ and $y$ if and only if $q \equiv 1 \pmod 4$ or $q=2$. 
\end{theorem}

In view of Theorem \ref{Genus}, not all  $q\times q$ square lattices with $q$ prime can be characterized by the quadratic form $x^{2}+y^{2}=q$, that is, not all $q\times q$ square lattices with $q$ being prime can be reproduced algebraically.

Via a combinatorial approach, it is possible to cover the cases which cannot be addressed algebraically, that is, the instances where $q$ cannot be represented as a sum of two squares. In this case, $q$ is represented by another quadratic form and consequently the code is realized on other lattices, such as the hexagonal lattice whose quadratic form is given by the norm in the ring of Eisenstein integers. However, since
\begin{equation}\label{eq00}
\frac{V(A \mathbb{Z}^{2})}{V(\mathbb{Z}^{2})} = |\det A|=q,
\end{equation}
\noindent we can still find a $2 \times 2$ matrix $A$ of determinant $q$ and realize the code as the set of cosets of $A \mathbb Z^2$ in $\mathbb Z^2$.  

Let $q=2n+1$, where $n\geq 2$, and consider the $q\times q$ square lattice. From the combinatorial viewpoint, the problem of determining the lattice vector $(x,y)\in \mathbb{Z}_{q}\times \mathbb{Z}_{q}$ which represents a polyomino is equivalent to finding a matrix 
\begin{equation}
\label{matrixA}
A=\left( \begin{array}{cc}
                                          c & d \\
                                          \alpha & \beta \\
                                        \end{array}
                                      \right)
\end{equation}
whose determinant equals $q$.  

The rows of $A$ define the moves on the square lattice, which in turn determine the representatives of the polyominoes (codewords) in the following way: $c$ units to the right and $d$ units down; $\alpha$ units to the right and $\beta$ units down. If any of the above integers is negative, then the move is in the opposite direction. This operation can also be seen as the modulo $q$-sum of vectors, where we consider the set of vectors generated by $(c,d)$ or by $(\alpha,\beta)$.

From there, it is concluded that the values of the first coordinate of the row vectors from the matrix $A$, that is, $c$ and $\alpha$, characterize the move in the horizontal direction on the square lattice and the values of $d$ and $\beta$, which determine the second coordinate of the corresponding row vectors, characterize the move in the vertical direction. 

In this way we obtain a set, denoted by $\mathcal{A}$, consisting of $q$ lattice vectors which correspond to the representatives of the polyominoes (codewords). If we change the values of $c$, $d$, $\alpha$ and $\beta$ so that $\det(A)=q$, the set of representatives of the polyominoes still has $q$ elements, however they can occupy different positions in the lattice. Hence, each set of positions yields a possibly different tessellation by  polyominoes. Despite the fact that the parameters of the toric quantum code do not depend on the shape of the polyomino, the positions of the representatives affect the minimum distance of the code.

As soon as the subspace $\mathcal{A}$ given by the representatives (codewords) is known, it is possible to choose the polyominoes that tessellate the $q\times q$ square lattice and, from there, the toric quantum code associated with this tessellation may be established in the same way as it was in the Kitaev's code.           

In \cite{ijam}, the authors provide the lattice vector $(1,g)$ which generates $q=2n+1$ lattice vectors that correspond to the representatives of the polyominoes (codewords) on the square lattice $q\times q$, where $g=2(n-1)$ and $n\in \mathbb{N}_{\geq 2}$. Moreover, for $g=2(n-1)$ not a multiple of 3, the corresponding code formed by the $q=2n+1$ representatives of the polyominoes (codewords) is a perfect code, that is, it has only one representative (codeword) $X$ in each row or column \cite{Sueli} and, for $g=2(n-1)$ a multiple of 3, the corresponding code formed by the $q=2n+1$ representatives of the polyominoes (codewords) is neither a perfect code \cite{Sueli} nor a quasi-perfect code \cite{quasiperf}.

Consequently, there are values of $q=2n+1$, with $q\not \equiv 1 \pmod 4$, where the respective code formed by $q=2n+1$ representatives of the polyominoes (codewords) on the square lattice $q\times q$ which are generated by the lattice vector $(1,g=2(n-1))$ is a classical perfect code in the sense of \cite{Sueli}. Therefore, it is possible to construct classical perfect codes on the square lattice $q\times q$ even if $q$ is not a sum of squares.

\subsection{Set of representatives of the polyominoes (codewords) on the $q\times q$ square lattice for odd $q \geq 5$}\label{codewords}

Consider the $q \times q$ square lattice, where $q=2n+1$ and $n\in \mathbb{N}_{\geq 2}$. The lattice vector $(1,g)$, where $g=2(n-1)$, generates $q$ lattice vectors which correspond to the representatives of the polyominoes (codewords) on the $q \times q$ square lattice \cite{ijam}. Moreover, when $n \not \equiv 1 \pmod{3}$, the corresponding code formed by the $q=2n+1$ representatives of the polyominoes (codewords) is a perfect code, that is, it has only one representative (codeword) $X$ in each row or column \cite{Sueli} and when $n \equiv 1 \pmod{3}$, the corresponding code formed by the $q=2n+1$ representatives of the polyominoes (codewords) is neither a perfect code \cite{Sueli} nor a quasi-perfect code \cite{quasiperf}.

If we set $\alpha$ and $\beta$ of matrix $A$ in \eqref{matrixA} equal to $1$ and $g=2(n-1)$, respectively, then from $\beta c - \alpha d = gc-d = q$ we can determine $c$ and $d$. Each solution, except that where either $c$ or $d$ is congruent to $0$ modulo $q$, generates the same set $\mathcal{A}$ of representatives, i.e., $\left\langle (1,g) \right\rangle = \left\langle (c,d) \right\rangle $ \cite{Clarice}. Thenceforward, we define $S$ as being the set of such solutions for $c$ and $d$, that is, $S$ is the set of all possible generators of the representatives of the polyominoes (codewords) on the $q\times q$ square lattice. 

Once the subspace formed by the representatives is known, then it is possible to choose the polyominoes which tessellate the square lattice and define the associated toric quantum code. The length of the constructed code is equal to the number of edges of the polyomino. Since the latter has area equal to $q$ and each edge belongs simultaneously to two square faces of the lattice on the torus, it follows that the effective number of edges is given by $n=2q$. The dimension of the code is given by $k=2$ (since the code is constructed on the torus) and the minimum distance of the code is given by the smallest distance between two distinct representatives of the polyominoes, i.e., $d_{M}=\min\{|x|+|y| \ | \ (x,y)\in \mathcal{A}\}$ and $w_{M}(x,y)=|x|+|y|$ is known as the \textit{Mannheim weight} of $(x,y)$. Thus, the parameters of the toric quantum code defined by the corresponding tessellation are $[[2q,2,d_{M}]]$. 

We emphasize that for each tessellation of the square lattice $\mathbb{Z}_{q} \times \mathbb{Z}_{q} = \mathbb{Z}_{q}^ {2}$, the polyominoes may have different shapes, however the parameters $n$, $k$, and $d_{M}$ do not change, i.e., the code that is obtained is the same. Nonetheless, the duality of the lattice on which the toric quantum code is constructed influences the error correction pattern \cite{LivroBombin}, that is, since the $\mathbb{Z}^{2}$-lattice is autodual, then the channel without memory is symmetric.

\subsubsection{Set of representatives of the polyominoes (codewords) on the $5\times 5$ square lattice}
Consider the $q\times q$ square lattice, where $q=5=2n+1$ (hence, $n=2$). The lattice vector $(1,g)$, where $g=2(n-1)=2$, generates the $q=5$ lattice vectors that correspond to the representatives of the polyominoes (codewords) on the $5\times 5$ square lattice. From $$A=\left( \begin{array}{cc}
                                          c & d \\
                                          1 & 2 \\
                                        \end{array}
                                      \right),$$
it follows that $\det(A)=2c-d=5=q$, that is, 
\begin{equation}
\label{dc}
d=2c-5. 
\end{equation}
All the solutions to the latter equation such that
\begin{equation}
\label{dcrange}
c,d\in \{-4,-3,-2,-1,1,2,3,4\}
\end{equation} 
yield the same set $\mathcal{A}$ of representatives, i.e., $\left\langle (1,2) \right\rangle = \left\langle (c,d) \right\rangle$ \cite{Clarice}. 

The five lattice vectors generated by $(1,2)$, which correspond to the representatives of the polyominoes (codewords) on the $5\times 5$ square lattice, are represented by X in Table \ref{fig1}. Table \ref{code5} shows the codewords derived from the representatives in Table \ref{fig1}.
{\small \begin{table}[h!] \caption{Representatives in the $5\times 5$ square lattice} \label{fig1}
\begin{center} \begin{tabular}{|c|c|c|c|c|c|}
\hline     &  0  &  1  &  2  &  3  &  4  \\
\hline   0 &  X  &     &     &     &     \\
\hline   1 &     &     &     &  X  &     \\
\hline   2 &     &  X  &     &     &     \\
\hline   3 &     &     &     &     &  X  \\
\hline   4 &     &     &  X  &     &     \\
\hline \end{tabular} \end{center} \end{table}}
{\small \begin{table}[h!] \caption{Codewords associated to the representatives in Table \ref{fig1}} \label{code5}
\begin{center} \begin{tabular}{|c|c|c|}
\hline                 &  \mbox{Codes}  &              \\
\hline  \mbox{Column}  &                &  \mbox{Row}  \\
\hline     00          &                &     00       \\
\hline     12          &                &     13       \\
\hline     24          &                &     21       \\
\hline     31          &                &     34       \\
\hline     43          &                &     42       \\
\hline \end{tabular} \end{center} \end{table}}

By inspection, one can now verify that the set $S$ of pairs $(c,d)$ that satisfy \eqref{dc}, \eqref{dcrange} is given by
\begin{eqnarray}
\label{geradores5}
S&=&\{(-4,-3), (-4,2), (-3,4), (-3,-1), (-2,1),(-2,-4), (-1,3),(-1,-2), \nonumber \\
&& \, \ (1,-3), (1,2), (2,-1),(2,4),(3,1),(3,-4),(4,3),(4,-2)\}                         
\end{eqnarray}
\noindent and $S$ is the set of all possible generators of the representatives of the polyominoes (codewords) on the $5\times 5$ square lattice which are given in Table \ref{fig1}.

\subsubsection{Set of representatives of the polyominoes (codewords) on the $q\times q$ square lattice}\label{codes}

Let $q=2n+1$ and $n\in \mathbb{N}_{\geq 2}$, then $\det(A)=gc-d=q$, i.e., $d=gc-q$. From the latter relation we can determine all possible values for $c$ and $d$. Each solution $(c,d)$, except that where either $c$ or $d$ is congruent to $0$ modulo $q$, generates the same set $\mathcal{A}$ of representatives. Hence, for $c, d$ belonging to $\{-(q-1),\ldots,-1,1,\ldots,q-1\}$, the next procedure can be used to determine all the pairs $(c,d)$.
\begin{center}
{\bf Procedure for determing all possible pairs $(c,d)$:}
\end{center}
{\bf Input}: $n\in \mathbb{N}_{\geq 2}$ .
\begin{enumerate}
\item $g:=2(n-1)$; \ $q:=2n+1$;  \ $S:=\{ \}$;
\item {\bf for} $c$  {\bf in} $\{-(q-1),\ldots,-1,1,\ldots,q-1\}$ {\bf do}  \\
 $\frac{}{}$ \hspace{0.2mm} $d:=gc \bmod{q}$; \\
 $\frac{}{}$ \hspace{0.2mm} $S:=S \cup \{ (c,d), (c, d-q) \}$; \\
{\bf end for};
\item {\bf return} $S$;
\end{enumerate}

The set $S$ obtained via the above procedure consists of all possible generators of the representatives of the polyominoes (codewords) on the $q\times q$ square lattice. Since $\det(A)=gc-d=q$ and the modulo $q$-sum is performed, note that the lattice vector $(1,g)\in S$. Table \ref{tableS} shows the corresponding set $S$ for the values $q=5,7 \ \mathrm{and} \ 9$.   

\begin{table}[h!]
\caption{Set $S$ for $q=5,7 \ \mathrm{and} \ 9$}
\label{tableS}
\begin{center}
\begin{tabular}{|l|l|lll}
\cline{1-2}
\textbf{$q$} & \textbf{Set $S$}                                 &  &  &  \\ \cline{1-2}
\textbf{5}             & \begin{tabular}[c]{@{}l@{}}\{(-6, 4), (-6, -3), (-5, 1), (-5,-6), ( -4,5), (-4,-2), \\ (-3,2), (-3,-5), (-2,6), (-2,-1), (-1, 3), (-1,-4), \\ (1,4), (1,-3), (2, 1), (2, -6), (3, 5), (3,-2), \\ (4,2), (4,-5), (5,6), (5,-1), (6,3), (6,-4)\}\end{tabular}                 &  &  &  \\ \cline{1-2}
\textbf{7}             & \begin{tabular}[c]{@{}l@{}}\{(-6, 4), (-6, -3), (-5, 1), (-5, -6), (-4, 5), (-4, -2),\\  (-3, 2), (-3, -5), (-2, 6), (-2, -1), (-1, 3), (-1, -4), \\ (1, 4), (1, -3), (2, 1), (2, -6), (3, 5), (3, -2), \\ (4, 2), (4, -5), (5, 6), (5, -1), (6, 3), (6, -4)\}\end{tabular} &  &  &  \\ \cline{1-2}
\textbf{9}             & \begin{tabular}[c]{@{}l@{}}\{(-8, 6), (-8, -3), (-7, 3), (-7, -6), (-5, 6), (-5, -3),  \\ (-4, 3), (-4, -6), (-2, 6), (-2, -3), (-1, 3), (-1, -6),\\ (1, 6), (1, -3), (2, 3), (2, -6), (4, 6), (4, -3), \\ (5, 3), (5, -6), (7, 6), (7, -3), (8, 3), (8, -6)\}\end{tabular} &  &  &  \\ \cline{1-2}
\end{tabular}
\end{center}
\end{table}

Consequently, via the combinatorial method, a classical cyclic code $\mathcal{A}$ on the $q \times q$ square lattice can be constructed for all $q=2n+1$ with $n \in \mathbb N_{\geq 2}$. Therefore, the method can address the cases that cannot be approached by the algebraic method, that is, those when $q$ cannot be represented as a sum of two squares. 

\subsection{Minimum distance of the code generated by $(1,g)$ on the $q\times q$ square lattice}\label{distance}                                                      

This section provides the minimum distance of the codes constructed in section \ref{codes} and, as mentioned in section \ref{codewords}, such a minimum distance is given by $d_{M}=\min\{|x|+|y| \ | \ (x,y)\in \mathcal{A}\}$, where $w_{M}(x,y)=|x|+|y|$ is known as the \textit{Mannheim weight} of $(x,y)$. 

Each codeword labeled by $X$ can move either in the vertical direction or in the horizontal direction to the closest codeword. Then for $n\geq 2$ the next lemma shows a generator of the code that gives the minimum number of cells to be moved in the vertical direction so that two codewords are closer together. 

\begin{lemma}\label{vertical}
For $n=2$ ($n\geq 3$), we have at least two (three) cells to be moved in the vertical direction so that two codewords are closer together. 
\end{lemma}
\begin{proof}
As before, let $$A=\left( \begin{array}{cc}
                                          c & d \\
                                          1 & g \\
                                        \end{array}
                                      \right).$$ For $c, d$ belonging to $\{-(q-1),\ldots,-1,1,\ldots,q-1\}$, if $\det(A)=gc-d$ is a multiple of $q$, then the lattice vector $(c,d)$ also generates the code on the $q\times q$ square lattice. The code is the set $\mathcal{A}$ of representatives of the corresponding polyominoes. Since we seek to determine the minimum number of cells to be moved in the vertical direction so that two codewords are closer together and $c\neq 0$ modulo $q$ (because the vector $(1,g)$ generates the code, then we can observe a single codeword in each column of the square lattice), then let entry $c$ of matrix $A$ be equal to 1. Hence, for $d=-3$, one has $$A=\left( \begin{array}{cc}
                                          1 & -3 \\
                                          1 & g \\
                                        \end{array}
                                      \right)$$ and $\det(A)=g+3=2n-2+3=2n+1=q$ for $n\geq 2$. Thus, the lattice vector $(1,-3)$ also generates the code for all $n\geq 2$. It follows that for all $n\geq 2$, we have three cells to be moved in the vertical direction between two codewords.   

For $n=2$ and  $q=2n+1=5$, the vector $(1,g)$ with $g=2(n-1)=2$ generates the $q=5$ lattice vectors which correspond to the codewords of the code on the $5 \times 5$ square lattice. Those five lattice vectors, which correspond to the representatives of the polyominoes on the $5\times 5$ square lattice, are represented by the label $X$ on Table \ref{fig1}. Hence, for $n=2$, we observe two cells to be moved in the vertical direction between two codewords and since the vectors $(1,-1)$, $(1,1)$, $(-1,-1)$ and $(-1,1)$ do not belong to the set $S$ in \eqref{geradores5}, that is, those vectors are not generators of the representatives of the polyominoes on the $5\times 5$ square lattice given in Table \ref{fig1}, we conclude that at least two cells must be moved in the vertical direction so that two codewords are closer together.

Observe that the lattice vector $(-1,-2)$ which corresponds to the moves in the opposite directions is also a generator of the corresponding code. 

For $n\geq 3$, using $(1,-3)$ as the generator, we have three cells to be moved in the vertical direction between two codewords. However, we will show that the latter quantity cannot be smaller than 3. In fact, let
\begin{itemize}
\item $\det \left( \begin{array}{cc}
                                          1 & 2 \\
                                          1 & g \\
                                        \end{array}
                                      \right)=g-2=2n-2-2=2n-4$;

\item $\det \left( \begin{array}{cr}
                                          1 & -2 \\
                                          1 & g \\
                                        \end{array}
                                      \right)=g+2=2n-2+2=2n$; 

\item $\det \left( \begin{array}{cc}
                                          1 & 1 \\
                                          1 & g \\
                                        \end{array}
                                      \right)=g-1=2n-2-1=2n-3$; 
                                      
\item $\det \left( \begin{array}{cr}
                                          1 & -1 \\
                                          1 & g \\
                                        \end{array}
                                      \right)=g+1=2n-2+1=2n-1$;
\item $\det \left( \begin{array}{cc}
                                          -1 & 2 \\
                                          1 & g \\
                                        \end{array}
                                      \right)=-g-2=-2n+2-2=-2n$;

\item $\det \left( \begin{array}{cr}
                                          -1 & -2 \\
                                          1 & g \\
                                        \end{array}
                                      \right)=-g+2=-2n+2+2=-2n+4$; 

\item $\det \left( \begin{array}{cc}
                                          -1 & 1 \\
                                          1 & g \\
                                        \end{array}
                                      \right)=-g-1=-2n+2-1=-2n+1$; 
                                      
\item $\det \left( \begin{array}{cr}
                                          -1 & -1 \\
                                          1 & g \\
                                        \end{array}
                                      \right)=-g+1=-2n+2+1=-2n+3$,
\end{itemize}
\noindent observe that the respective determinants are different from zero and their corresponding absolute value is less than $q$ for $n\geq 3$, then for $n\geq 3$ the lattice vectors $(1,1)$, $(1,-1)$, $(1,2)$, $(1,-2)$, $(-1,1)$, $(-1,-1)$, $(-1,2)$, $(-1,-2)$ are not generators of the code on the $q \times q$ square lattice and so we conclude that, for $n\geq 3$, at least 3 cells must be moved in the vertical direction so that two codewords are closer together. Furthermore, for $n\geq 3$, we have $d_{M}(1,g)>d_{M}(1,-3)$. 

Observe that the lattice vector $(-1,3)$ which corresponds to the moves in the opposite directions is also a generator of the corresponding code. 
\end{proof}   

Now for $n\geq 3$ we determine a lattice vector which is different from $(1,-3)$ and $(-1,3)$, and gives the minimum number of cells to be moved in the horizontal direction so that two codewords are closer together.

\begin{lemma}\label{horizontal}
Let $g=3q'+r$ with $0\leq r \leq 2$ and $g=2n-2$. For $n\geq 3$ we have that $(q'+1)\neq 1$ is the minimum number of cells to be moved in the horizontal direction so that two codewords are closer together.
\end{lemma}
\begin{proof}
For $n\geq 2$ one has $q=2n+1$, $g=2n-2$ and $(1,g)$ which is one of the lattice vectors that generate the $q$ lattice vectors that correspond to the codewords from the code on the $q\times q$ square lattice.

Hence, using the vector $(1,g)$, we have that the move on the $q\times q$ square lattice is specified as follows: 1 unit to the right and $g$ units down. Consequently, there will be a cell labeled by $X$ located in the second column and $(g+1)$th row of the $q\times q$ square lattice. We now have $g$ unlabeled cells above the $X$ in the second column. 

According to Lemma \ref{vertical}, if we use the generator $(1,-3)$, then for $n\geq 3$ we have three cells between two codewords in the vertical direction. Therefore, write $g=3q'+r$ with $0\leq r \leq 2$. Since there are $g$ unlabeled cells above the $X$ in the second column, then the lattice vector $(q'+1,r)$ provides the minimum number of cells to be moved in the horizontal direction so that two codewords are closer together, consequently, $(q'+1)\neq 1$ is the minimum number of cells to be moved in the horizontal direction so that two codewords are closer together.  

Observe that the lattice vector $(-(q'+1),-r)$, which corresponds to the moves in the opposite directions, can also be used to provide the minimum number of cells to be moved in the horizontal direction so that two codewords are closer together. 
\end{proof}

Next we provide some examples to illustrate the minimum distance of the code on the square lattice.

\begin{example}\label{examplo3}

Let $n=3$. In Table \ref{tabela7} each label $X$ represents a codeword in the $7\times 7$ square lattice. The lattice vector $(1,g)=(1,2n-2)=(1,4)$ generates the corresponding code. Let $g=2n-2=6-2=4$. Since $g=3\cdot 1+1$, we have $q'=1$, $r=1$ and the corresponding vectors $(1,-3)$, $(-1,3)$, $(q'+1,r)=(2,1)$ and $(-2,-1)$. As one can observe on Table \ref{tabela7}, the corresponding minimum distance of this code is given by $d_{M}=\min\{|2|+|1|,|1|+|-3| \}=\min\{3,4\}=3$. 

{\small \begin{table}[h!] \caption{Codewords (representatives) in the $7\times 7$ square lattice} \label{tabela7}
\begin{center} \begin{tabular}{|l|c|c|c|c|c|c|c|}
\hline     &  0  &  1  &  2  &  3  &  4  &  5  &  6   \\
\hline   0 &  X  &     &     &     &     &     &      \\
\hline   1 &     &     &  X  &     &     &     &      \\
\hline   2 &     &     &     &     &  X  &     &      \\
\hline   3 &     &     &     &     &     &     &  X   \\
\hline   4 &     &  X  &     &     &     &     &      \\
\hline   5 &     &     &     &  X  &     &     &      \\
\hline   6 &     &     &     &     &     &  X  &      \\
 \hline \end{tabular} \end{center} \end{table}}
\end{example}

\begin{example}\label{examplo4}

Let $n=4$. In Table \ref{tabela9}, each label $X$ represents a codeword in the $9\times 9$ square lattice. The lattice vector $(1,g)=(1,2n-2)=(1,6)$ generates the corresponding code. Let $g=2n-2=8-2=6$. Since $g=3\cdot 2+0$, we have $q'=2$, $r=0$ and the corresponding vectors $(1,-3)$, $(-1,3)$, $(q'+1,r)=(3,0)$ and $(-3,0)$. As one can observe on Table \ref{tabela9}, the corresponding minimum distance of this code is given by $d_{M}=\min\{|3|+|0|,|1|+|-3| \}=\min\{3,4\}=3$.    
 
{\small \begin{table}[h!] \caption{Codewords (representatives) in the $9\times 9$ square lattice} \label{tabela9}
\begin{center} \begin{tabular}{|l|c|c|c|c|c|c|c|c|c|}
\hline     &  0  &  1  &  2  &  3  &  4  &  5  &  6  &  7  &  8  \\
\hline  0  &  X  &     &     &  X  &     &     &  X  &     &     \\
\hline  1  &     &     &     &     &     &     &     &     &     \\
\hline  2  &     &     &     &     &     &     &     &     &     \\
\hline  3  &     &     &  X  &     &     &  X  &     &     &  X  \\
\hline  4  &     &     &     &     &     &     &     &     &     \\
\hline  5  &     &     &     &     &     &     &     &     &     \\
\hline  6  &     &  X  &     &     &  X  &     &     &  X  &     \\
\hline  7  &     &     &     &     &     &     &     &     &     \\
\hline  8  &     &     &     &     &     &     &     &     &     \\
 \hline \end{tabular} \end{center} \end{table}} 
\end{example}

\begin{example}\label{examplo5}

Let $n=5$. In Table \ref{tabela11}, each label $X$ represents a codeword in the $11\times 11$ square lattice. The lattice vector $(1,g)=(1,2n-2)=(1,8)$ generates the corresponding code. Let $g=2n-2=10-2=8$. Since $g=3\cdot 2+2$, we have $q'=2$, $r=2$ and the corresponding vectors $(1,-3)$, $(-1,3)$, $(q'+1,r)=(3,2)$ and $(-3,-2)$. As one can observe on Table \ref{tabela11}, the corresponding minimum distance of this code is given by $d_{M}=\min\{|3|+|2|,|1|+|-3| \}=\min\{5,4\}=4$. 
 
{\small \begin{table}[h!]\caption{Codewords (representatives) in the $11\times 11$ square lattice} \label{tabela11} 
\begin{center} \begin{tabular}{|l|c|c|c|c|c|c|c|c|c|c|c|}
\hline     &  0  &  1  &  2  &  3  &  4  &  5  &  6  &  7  &  8  &  9  &  10  \\
\hline  0  &  X  &     &     &     &     &     &     &     &     &     &      \\
\hline  1  &     &     &     &     &     &     &     &  X  &     &     &      \\
\hline  2  &     &     &     &  X  &     &     &     &     &     &     &      \\
\hline  3  &     &     &     &     &     &     &     &     &     &     &   X  \\
\hline  4  &     &     &     &     &     &     &  X  &     &     &     &      \\
\hline  5  &     &     &  X  &     &     &     &     &     &     &     &      \\
\hline  6  &     &     &     &     &     &     &     &     &     &  X  &      \\
\hline  7  &     &     &     &     &     &  X  &     &     &     &     &      \\
\hline  8  &     &  X  &     &     &     &     &     &     &     &     &      \\
\hline  9  &     &     &     &     &     &     &     &     &  X  &     &      \\
\hline  10 &     &     &     &     &  X  &     &     &     &     &     &      \\
 \hline \end{tabular} \end{center} \end{table}}
\end{example}   

\begin{example}\label{examplo6}

Let $n=6$. In Table \ref{tabela13}, each label $X$ represents a codeword in the $13\times 13$ square lattice. The lattice vector $(1,g)=(1,2n-2)=(1,10)$ generates the corresponding code. Let $g=2n-2=12-2=10$. Since $g=3\cdot 3+1$, we have $q'=3$, $r=1$ and the corresponding vectors $(1,-3)$, $(-1,3)$, $(q'+1,r)=(4,1)$ and $(-4,-1)$. As one can observe on Table \ref{tabela13}, the corresponding minimum distance of this code is given by $d_{M}=\min\{|4|+|1|,|1|+|-3| \}=\min\{5,4\}=4$.
 
{\small \begin{table}[h!]\caption{Codewords (representatives) in the $13\times 13$ square lattice} \label{tabela13} 
\begin{center} \begin{tabular}{|l|c|c|c|c|c|c|c|c|c|c|c|c|c|}
\hline     &  0  &  1  &  2  &  3  &  4  &  5  &  6  &  7  &  8  &  9  &  10  &  11 &  12  \\
\hline  0  &  X  &     &     &     &     &     &     &     &     &     &      &     &      \\
\hline  1  &     &     &     &     &  X  &     &     &     &     &     &      &     &      \\
\hline  2  &     &     &     &     &     &     &     &     &  X  &     &      &     &      \\
\hline  3  &     &     &     &     &     &     &     &     &     &     &      &     &   X  \\
\hline  4  &     &     &     &  X  &     &     &     &     &     &     &      &     &      \\
\hline  5  &     &     &     &     &     &     &     &  X  &     &     &      &     &      \\
\hline  6  &     &     &     &     &     &     &     &     &     &     &      &  X  &      \\
\hline  7  &     &     &  X  &     &     &     &     &     &     &     &      &     &      \\
\hline  8  &     &     &     &     &     &     &  X  &     &     &     &      &     &      \\
\hline  9  &     &     &     &     &     &     &     &     &     &     &   X  &     &      \\
\hline  10 &     &  X  &     &     &     &     &     &     &     &     &      &     &      \\
\hline  11 &     &     &     &     &     &  X  &     &     &     &     &      &     &      \\
\hline  12 &     &     &     &     &     &     &     &     &     &  X  &      &     &      \\
 \hline \end{tabular} \end{center} \end{table}} 

\end{example}   

Therefore, by using the lattice vectors $(1,-3)$, $(-1,3)$, $(q'+1,r)$ and $(-(q'+1),-r)$, the next theorem guarantees that the minimum Mannheim distance $d_{M}$ of the code generated by the vector $(1,g)$ is 4, where $n\geq 5$.  

\begin{theorem}\label{distmin}       
Let $n\geq 5$, $g=2n-2$ and $q=2n+1$. The minimum Mannheim distance $d_{M}$ of the code generated by $(1,g)$ on the $q\times q$ square lattice equals 4.
\end{theorem}
\begin{proof}
For $n=5$ and $n=6$, it follows from Examples \ref{examplo5} and \ref{examplo6}, respectively, that the minimum Mannheim distance $d_{M}$ of the  corresponding codes is given by $d_{M}=4$. The statement of the theorem is now proved by induction on $n$. Let $n\geq 5$.  Then $g=2n-2=2(n-1)=3q'+r$ with $0\leq r\leq 2$, $q=2n+1$ and $(1,g)$ generates the code on the $q\times q$ square lattice. Since the moves from one cell to another are restricted to either the vertical direction or the horizontal direction, the minimum Mannheim distance $d_{M}$ of the code generated by $(1,g)$ on the $q\times q$ square lattice is given by $d_{M}=\min\{|q'+1|+|r|,|1|+|-3|\}$.

By induction hypothesis, suppose 
\begin{equation}
\label{dM}
d_{M}=\min\{|q'+1|+|r|,|1|+|-3|\}=4
\end{equation}
for $n\geq 5$. Then
\begin{enumerate}
\item If $r=0$, then $|q'+1|\geq 4$;    
\item If $r=1$, then $|q'+1|\geq 3$;
\item If $r=2$, then $|q'+1|\geq 2$.  
\end{enumerate}
We now show that the claim in \eqref{dM} holds for $n+1$. In this case, $g=2(n+1)-2=2n+2-2=(2n-2)+2=(3q'+r)+2=3q'+(r+2)$. From the induction hypothesis, we have:
\begin{enumerate}
\item If $r=0$, then $|q'+1|\geq 4$ and $g=3q'+(r+2)=3q'+2$. Therefore $d_{M}=min\{|q'+1|+2,|1|+|-3|\}=4$, since $|q'+1|+2\geq 6$; 

\item If $r=1$, then $|q'+1|\geq 3$ and $g=3(q'+1)+0$. Therefore $d_{M}=min\{|q'+2|+0,|1|+|-3|\}=min\{|q'+1|+1,|1|+|-3|\}=4$, since $|q'+1|+1\geq 4$;  

\item If $r=2$, then $|q'+1|\geq 2$ and $g=3(q'+1)+1$. Therefore $d_{M}=min\{|q'+2|+1,|1|+|-3|\}=min\{|q'+1|+2,|1|+|-3|\}=4$, since $|q'+1|+2\geq 4$.          
\end{enumerate}
In conclusion, the minimum Mannheim distance $d_{M}$ equals 4 for any $n \geq 5$.
\end{proof}

From Theorem \ref{distmin}, the parameters of the corresponding toric quantum codes are $[[2q,2,d_{M}=3]]$ for $n=2,3,4$, and $[[2q,2,d_{M}=4]]$ for $n\geq 5$.

\subsection{Tessellation on the $q\times q$ square lattice by using polyomino}\label{Polyomino}

For $n=2$, then we have $q=5$. The square lattice $5\times 5$ can be tessellated by using two shapes for the polyomino, one of them is the Lee sphere of radius 1 and the other one can be considered as the union of a square of size $2\times 2$ with a square of size $1\times 1$. Figure 3 illustrates the corresponding tessellations on the $5\times 5$ square lattice by using these two polyomino shapes.

\begin{figure}[h!]
    \centering
    \begin{minipage}{0.49\textwidth}
        \centering
        \includegraphics[scale=0.48]{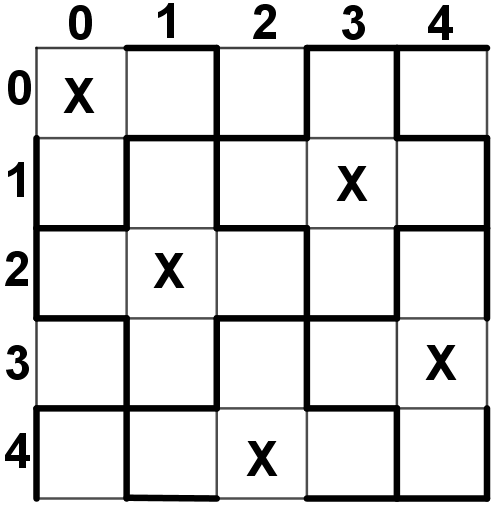} 
    \end{minipage} 
    \begin{minipage}{0.49\textwidth}
        \centering
        \includegraphics[scale=0.51]{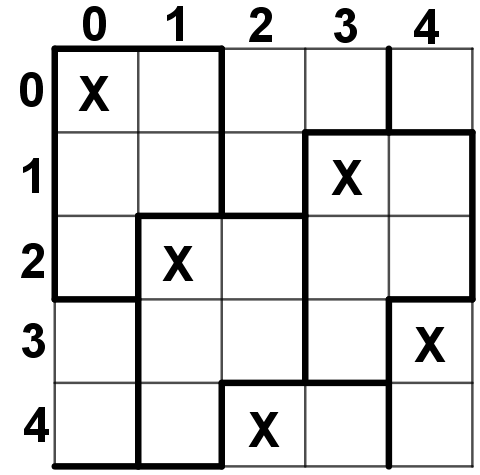} 
    \end{minipage}
    \caption{Two representatives of regions with area 5.}
\end{figure}

\begin{figure}[h!]
    \centering
    \begin{minipage}{0.49\textwidth}
        \centering
        \includegraphics[scale=0.5]{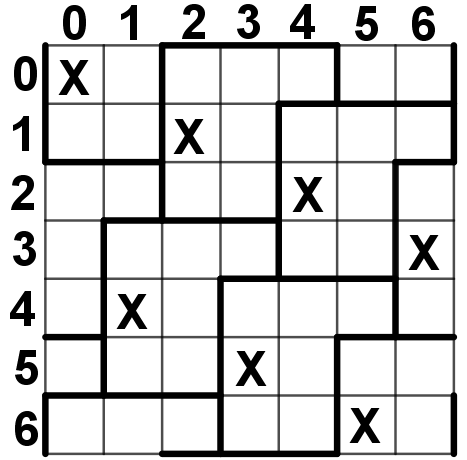} 
    \end{minipage} 
    \begin{minipage}{0.49\textwidth}
        \centering
        \includegraphics[scale=0.5]{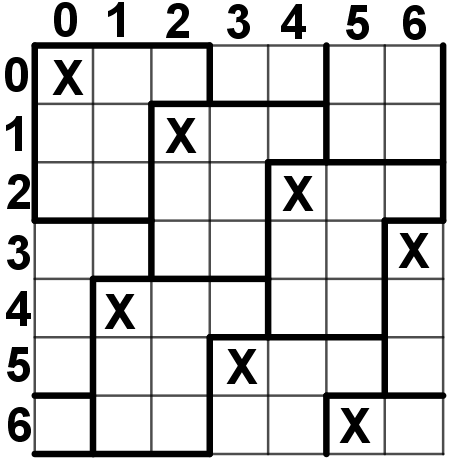} 
    \end{minipage}
    \caption{Two representatives of regions with area 7.}
    \label{fig7x7}
\end{figure}

\begin{figure}[ht!]
    \centering
    \begin{minipage}{0.49\textwidth}
        \centering
        \includegraphics[scale=0.5]{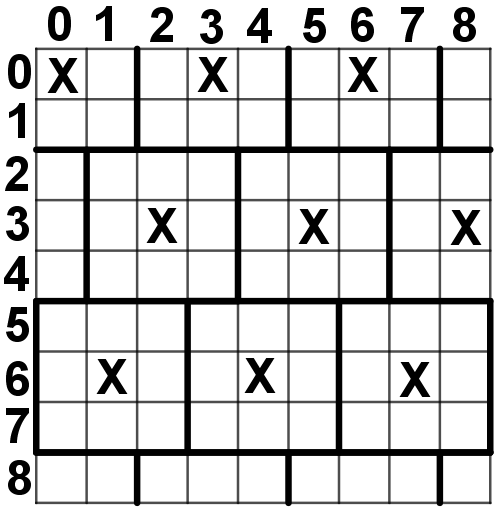} 
    \end{minipage} 
    \begin{minipage}{0.49\textwidth}
        \centering
        \includegraphics[scale=0.5]{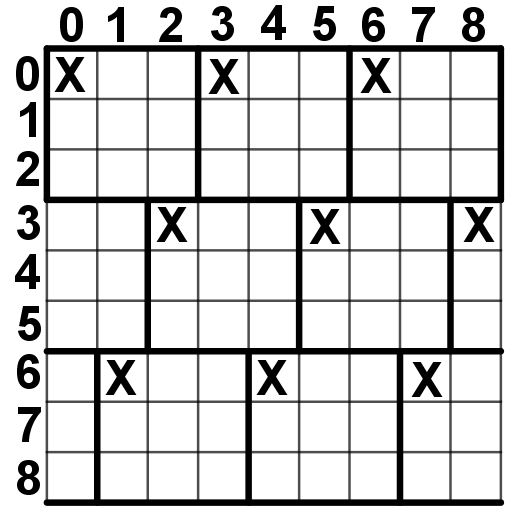} 
    \end{minipage}
    \caption{Two representatives of regions with area 9.}
    \label{fig9x9}
\end{figure}

\begin{figure}[ht!]
    \centering
    \includegraphics[scale=0.6]{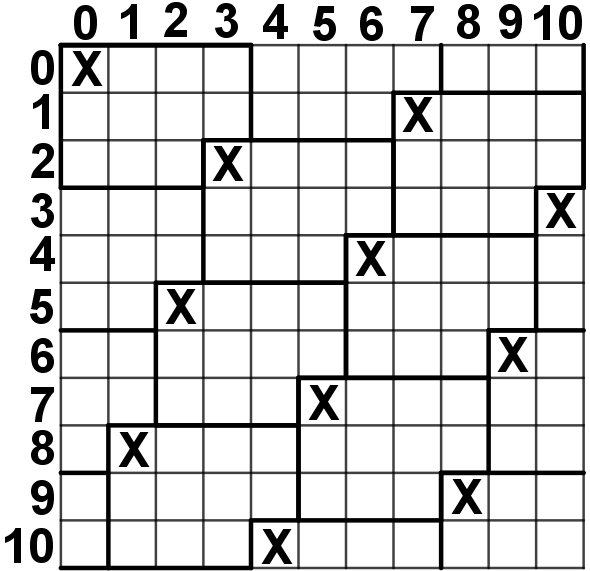}
    \caption{Tessellation on the square lattice $11\times 11$.}
    \label{fig11x11}
\end{figure}

\begin{figure}[ht!]
    \centering
    \includegraphics[scale=0.7]{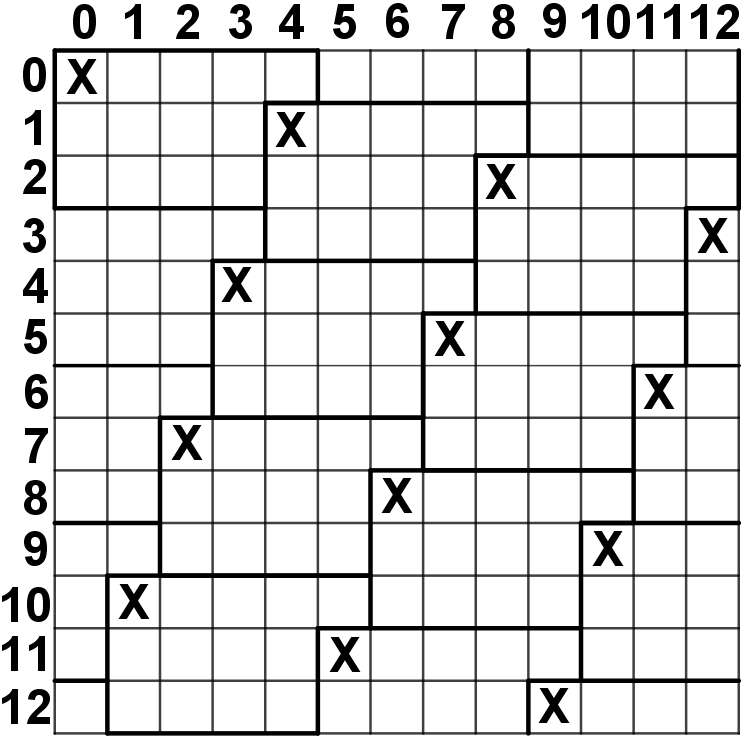}
    \caption{Tessellation on the square lattice $13\times 13$.}
    \label{fig13x13}
\end{figure}

\begin{figure}[ht!]
    \centering
    \includegraphics[scale=0.6]{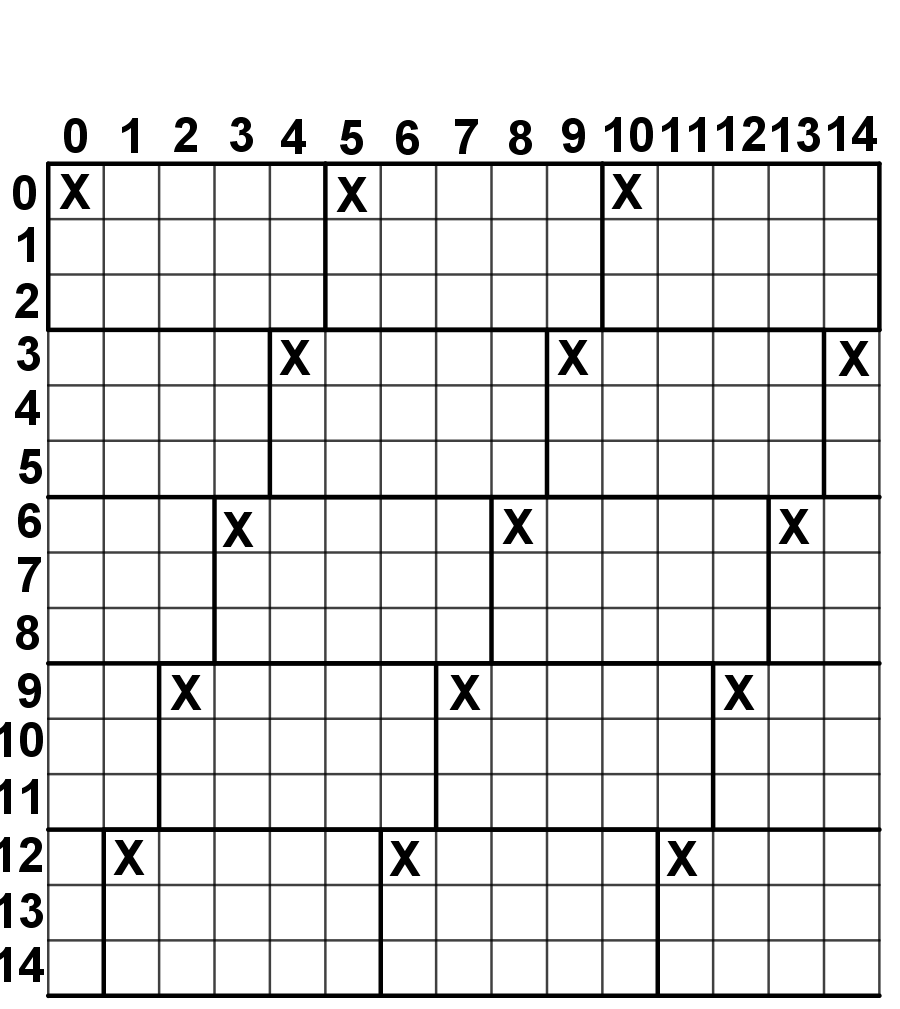}
    \caption{Tessellation on the square lattice $15\times 15$.}
    \label{fig15x15}
\end{figure}

Let $n\geq 3$ and $q=2n+1$. Lemmas \ref{vertical} and \ref{horizontal} supply a systematic way to obtain the polyomino shape that can tessellate the $q\times q$ square lattice by considering the polyomino shape as the union of a rectangle of size $(q'+1)\times 3$ with a rectangle of size $1\times r$. Observe that every square is also a rectangle. Figures \ref{fig7x7}, \ref{fig9x9}, \ref{fig11x11}, \ref{fig13x13} and \ref{fig15x15} illustrate the corresponding tessellations on the $7\times 7$, $9\times 9$, $11\times 11$, $13\times 13$ and $15\times 15$ square lattices, respectively.  


Notice that these polyomino shapes have exactly $q$ square faces (cells). In fact, let $g=2n-2=3q'+r$ where $n\geq 3$, then $r=0,1,2$. If $r=0$, then $q'+1=\frac{2n+1}{3}=\frac{q}{3}$. Since, in this case, the polyomino shape is a rectangle of size $(q'+1)\times 3$, then the amount of square faces is $\frac{q}{3}\cdot 3=q$. Now if $r=1$, then $q'+1=\frac{2n}{3}$. Since, in this case, the polyomino shape is the union of a rectangle of size $(q'+1)\times 3$ with a rectangle of size $1\times r$, then the amount of square faces is $(\frac{2n}{3}\cdot 3)+(1\cdot 1)=2n+1=q$. Finally if $r=2$, then $q'+1=\frac{2n-1}{3}$. Since, in this case, the polyomino shape is the union of a rectangle of size $(q'+1)\times 3$ with a rectangle of size $1\times r$, then the amount of square faces is $(\frac{2n-1}{3}\cdot 3)+(1\cdot 2)=2n-1+2=2n+1=q$. 

Therefore through a systematic way it is provided in this section the polyomino shapes that can tessellate the square lattice $q\times q$, where $q=2n+1$ and $n\geq 2$. As mentioned before, the channel without memory to be considered is symmetric, since the $\mathbb{Z}^{2}$-lattice is autodual \cite{LivroBombin}.

\section{Interleaved Toric Quantum Codes}

The parameters of the toric quantum codes constructed in section \ref{toricquantum} are given by $[[2q,k=2,d_{M}=3]]$ ($q=2n+1$), for $n=2,3,4$, and $[[2q,k=2,d_{M}=4]]$ ($q=2n+1$), for $n\geq 5$. A toric quantum code with minimum distance $d_{M}$ is able to correct up to $t$ errors, where $t=\lfloor \frac{d_{M}-1}{2} \rfloor$ \cite{Shor}, therefore, such codes from section \ref{toricquantum} are able to correct up to $t=1$ error.  

Section \ref{Polyomino} provides the polyomino shapes that can tessellate the $q\times q$ square lattice. 

It is shown in this section that the combination of the corresponding toric quantum codes with parameters $[[2q,k=2,t=1]]$ and the interleaving results in interleaved toric quantum codes with parameters $[[2q^{2},2q,t_{i}=q]]$, where $t_{i}$ is the interleaved toric quantum code error correcting capability.      

References \cite{ijam} and \cite{celso} provide effective interleaving techniques for combating burst of errors by using classical codes. However, to the best of our knowledge, little is known regarding interleaving techniques for combating cluster of errors in toric quantum codes. 

We assume that the clusters of errors have a polyomino shape which is given in section \ref{Polyomino} and the qubits are in a one-to-one correspondence with the edges of the $q\times q$ square lattice.

\begin{figure}[h!]
   \centering
   \includegraphics[scale=0.6]{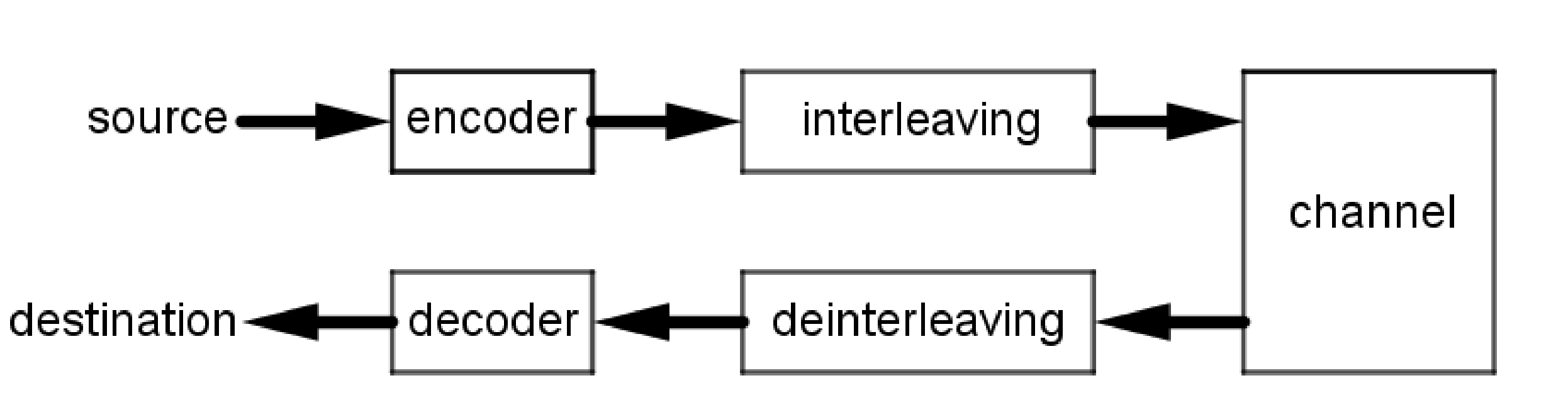}
   \caption{Model of an storage system, from \cite{celso}.}
   \label{fig9}
  \end{figure}

Figure \ref{fig9} shows the model of the storage system under consideration.

Let the $q\times q$ square lattice, where $q\geq 5$. Such a square lattice has a total of $2q^{2}$ qubits (edges). From now on we explain how to spread $2q^{2}$ adjacent qubits on the $q\times q$ square lattice. Observe that the toric quantum code constructed in section \ref{toricquantum} on such a square lattice consists of $q$ codewords whose fundamental region \cite{ClariceArtigo} is a polyomino which has $q$ square faces and $2q$ edges. 

Thenceforth the first $2q$ adjacent edges (qubits) are arranged as it follows: the first $q$ edges are placed on the square faces (cells) related to the codewords ($X$ marks); notice that it is needed to follow the sequence of the codewords by using the generator lattice vector $(1,g)$ and initiating with the null lattice vector $(0,0)$. The other $q$ edges are placed in the same way on the same square faces. 

Now the other $q-1$ blocks with $2q$ adjacent edges (qubits) are spread across the square lattice analogously, that is, suppose that we wish to separate another block with $2q$ adjacent edges (qubits) on the square lattice. As the fundamental region is a polyomino which has $q$ square faces including the square face related to the codeword ($X$ mark), then we start the corresponding arrangement putting the first edge of the $q$ first edges in one of the $q-1$ square faces related to the codeword $(0,0)$ and use the generator lattice vector $(1,g)$ to follow the sequence of the codewords to spread the other $q-1$ edges in the corresponding square faces. The other $q$ adjacent edges are placed in the same way on the same square faces. 


Note that by spreading every $q$ adjacent edges from the block of $2q$ adjacent edges as previously mentioned, we have each one of these $q$ edges belonging to a different fundamental region, that is, each one of these $q$ edges belongs to a different codeword.

Since the toric quantum code constructed on the $q\times q$ square lattice in section \ref{toricquantum} has $q$ codewords and it is able to correct up to $t=1$ error, then by using the interleaving method previously described it is possible to correct up to $q$ errors in burst from the total of $2q^{2}$ qubits. In fact,  assume that a cluster of $q$ errors in the channel has a polyomino shape which is given in section \ref{Polyomino}. Therefore, when the deinterleaving process is applied, each one of these $q$ errors occurs, respectively, in one of the $q$ codewords of the toric quantum code $[[2q,k=2,t=1]]$ and then such a code is applied to correct these $q$ errors in burst.    

Therefore, in this section, it is shown that the combination of the corresponding toric quantum code with parameters $[[2q,k=2,t=1]]$ and the presented interleaving method results in an interleaved toric quantum code with parameters $[[2q^{2},2q,t_{i}=q]]$, where $t_{i}$ is the interleaved toric quantum code error correcting capability.

The code rate \cite{rates} and the coding gain \cite{rates} are given by $R=\dfrac{k}{n}$ and $G=\dfrac{k}{n}(t+1)$, respectively, where $t$ is the toric quantum code error correcting capability. Consequently, the code rate and the coding gain of the interleaved toric quantum code are given, respectively, by $R_{i}=\dfrac{1}{q}$ and $G_{i}=\dfrac{q+1}{q}=1+\dfrac{1}{q}$, where $q=2n+1$ and $n\geq 2$.

The parameters of Kitaev's toric quantum code are given by $[[2q^{2},2,t_{k}=\lfloor \dfrac{q-1}{2} \rfloor]]$, where $t_{k}$ is the Kitaev's toric quantum code error correcting capability and $q=2n+1$, with  $n\geq 2$. Therefore the code rate and the coding gain of Kitaev's toric quantum code are given, respectively, by $R_{k}=\dfrac{1}{q^{2}}$ and $G_{k}=\dfrac{1}{q^{2}}(t_{k}+1)$. Since $\dfrac{1}{q^{2}} < \dfrac{1}{q}$ and $\lfloor \dfrac{q-1}{2} \rfloor < q+1$, then $R_{k} < R_{i}$ and $G_{k} < G_{i}$. Thenceforth the code rate and the coding gain of the interleaved toric quantum code are better than the code rate and the coding gain of Kitaev's toric quantum code for $q=2n+1$, where $n\geq 2$. 

In \cite{ClariceArtigo} the authors reproduce Bombin and Martin-Delgado's code. Let $m=2r^{2}+2r+1$, for $r=1,2,3,\ldots$. The parameters of Bombin and Martin-Delgado's toric quantum code are given by $[[2m,2,t_{bm}=r]]$, where $t_{bm}$ is the Bombin and Martin-Delgado's toric quantum code error correcting capability and $d_{bm}=2r+1$ is the code minimum distance. Let $q=2n+1$, where $n\geq 2$, and $r=q$, then the parameters of the corresponding toric quantum code are $[[2(2q^{2}+2q+1),2,t_{bm}=q]]$ and the code rate and the coding gain of this code are given, respectively, by $R_{bm}=\dfrac{1}{2q^{2}+2q+1}$ and $G_{bm}=\dfrac{1}{2q^{2}+2q+1}(q+1)$. Since $\dfrac{1}{2q^{2}+2q+1} < \dfrac{1}{q}$, then $R_{bm} < R_{i}$ and $G_{bm} < G_{i}$. Consequently, the code rate and the coding gain of the interleaved toric quantum code are better than the code rate and the coding gain of Bombin and Martin-Delgado's toric quantum code for $r=q=2n+1$, where $n\geq 2$.   

Table \ref{tablecodes} shows some equivalent interleaved toric quantum codes and their corresponding interleaving coding gains in dB when the cluster of errors has a polyomino shape on the $q\times q$ square lattice.

\begin{table}[h!] \caption{Array order, interleaved toric quantum code $[[2q^{2},2q,t_{i}=q]]$ and interleaving coding gain $G$} \label{tablecodes}
\begin{center} \begin{tabular}{|c|c|l|} \hline
\hline  $q$   &  $[[2q^{2},2q,t_{i}=q]]$  &  $G$, \mbox{dB} \\ \hline
\hline  5     &  $[[50,10,t_{i}=5]]$   &   1,2    \\
\hline  7     &  $[[98,14,t_{i}=7]]$  &  1,14286 \\
\hline  9     &  $[[162,18,t_{i}=9]]$  &  1,11111 \\
\hline  11    &  $[[242,22,t_{i}=11]]$    &  1,09091 \\
\hline  13    &  $[[338,26,t_{i}=13]]$  &  1,07692 \\
\hline  15    &  $[[450,30,t_{i}=15]]$ &  1,06667 \\
\hline  17    &  $[[578,34,t_{i}=17]]$ &  1,05882 \\
\hline \end{tabular} \end{center} \end{table}

Notice that $G$ converges to 1 as $q$ tends to infinity, consequently, the loss of interleaving coding gain is relatively small for large $q$ values.

\begin{remark}
The separation of qubits can be accomplished by using the transformation matrix $A=\left( \begin{array}{cc}
                                          c & d \\
                                          1 & g \\
                                        \end{array}
                                      \right)$ where in section \ref{toricquantum} all possible values for $c$ and $d$ are determined.  
\end{remark}

\section*{Acknowledgment}
The authors would like to thank the financial Brazilian agency FAPESP (Funda\c{c}\~{a}o de Amparo \`{a} Pesquisa do Estado de S\~{a}o Paulo), under grant no. 2013/03976-9, for the funding support.


\end{document}